\documentclass[11pt,a4paper]{article}
\usepackage{epsf,feynmf,amsmath}
\usepackage{graphicx,color,mathrsfs,amsmath,amssymb}
\title{\bf The renormalization group equations revisited}
\author{Jean-Fran\c cois Mathiot\thanks{e-mail: jean-francois.mathiot@clermont.in2p3.fr},\\
{\small \em  Universit\'e Clermont Auvergne, Laboratoire de Physique de Clermont,
} \\ {\small \em CNRS/IN2P3, BP10448, F-63000 Clermont-Ferrand, France}  }

\date{}

\begin{document}

\maketitle
\bibliographystyle{unsrt}
\abstract{Starting from a well defined local Lagrangian, we analyze the renormalization group equations in terms of the two different arbitrary scales associated with the regularization procedure and with the physical renormalization of the bare parameters, respectively. We apply our formalism to the minimal subtraction scheme using dimensional regularization. We first argue that the relevant regularization scale in this case should be dimensionless. By relating bare and renormalized parameters to physical observables, we calculate the coefficients of the  renormalization group equation up to two loop order in the $\phi^4$ theory. We show that the usual assumption, considering the  bare parameters to be independent of the regularization scale, is not a direct consequence of any physical argument. The coefficients that we find in our two-loop calculation are identical to the standard practice. We finally comment on  the decoupling properties of the renormalized coupling constant.}
\vspace{0.5cm}
\section{Introduction}
With the discovery of a scalar particle with a mass of about $125$ MeV, which is by now identified as the so-called Higgs boson \cite{higgstheo}, particle physics has entered into a completely new area. This discovery has  led to numerous discussions as far as the scale of new physics is concerned. 
In order to have an unambiguous hint on this scale, either from experimental measurements or from theoretical considerations, it is however essential to  clearly identify the relevant scales inherent to any theoretical calculation of a physical observable and specify their intrinsic nature. Among them, we shall disantangle in this study the scale associated to the regularization procedure from the one associated to the physical renormalization of the bare parameters. These are often mixed in practical calculations. As we shall argue in this study, the relevant regularization scale in dimensional regularization ($DR$) should be {\it dimensionless}. It is also not a physical quantity in the sense that it cannot be determined by any experimental measurement. This is in contrast to what we shall call the renormalization point which is a measured, {\it dimensionful},  physical quantity.

Once this is done, one should rely on well-defined and universal hypothesis in order to be able to extract meaningful physical patterns. This should be done by always connecting nonphysical quantities to physical observables. These physical observables, either calculated theoretically and/or measured experimentally, should be the only one on which one should rely in order to make theoretical hypothesis. This will be our guiding principle in order to identify the behavior of the parameters of the theory, like coupling constants and masses, as a function of the regularization scale or renormalization point. This is indeed what is already done for the derivation of the renormalization group ($RG$) equations from the invariance of the $S$ matrix \cite{stuec,collins}. This is however never done in a similar way  for the calculation of their coefficients $\beta, \gamma_m$ or $\gamma$. 

Finally, one should rely on a well defined Lagrangian which respects the fundamental symmetries  we attribute to Nature, and in particular gauge and Poincar\'e symmetries. We shall consider in our study {\it a local Lagrangian  constructed from local products of field operators or derivatives of field operators}. This has the advantage to avoid the introduction of any {\it had-hoc} mass scale - like a finite cut-off in momentum space used for regularization purposes - which will inevitably pollute the identification of the relevant physical scale at which new physics should show up. The only relevant mass scale one should consider in this respect appears in the general expansion of the  Lagrangian in terms of effective operators ${\cal O}_{ni}$
\begin{equation} \label{Leff}
{\cal L}={\cal L}_{SM}+\sum_{(n,i)\ge1} \frac{c_{ni}\ {\cal O}_{ni}}{\Lambda_{NP}^{n}},
\end{equation}
where ${\cal L}_{SM}$ is the Lagrangian of the Standard Model. 
In the absence of knowledge of the most general Lagrangian  ${\cal L}$, {\it i.e.} in the absence of knowledge of the  value of the coefficients $c_{ni}$, the scale $\Lambda_{NP}$ above which new physics becomes sizeable is unknown. 
This local Lagrangian should be well-defined, to start with, from a mathematical point of view. This can be done for instance using the recently proposed Taylor-Lagrange regularization scheme ($TLRS$) which properly takes into account the nature of quantum fields as operator valued distributions \cite{GW, GW_gauge}. This is also done, as is well known, with $DR$  in a space-time of dimension $D=4-\epsilon$. We shall consider in this study this latter regularization procedure in order to compare our results with  present calculations in the Standard Model.

The plan of the article is as follows. We discuss in Sec.~\ref{general} the general spirit of our study, with the identification of the relevant scales and the relevant coupling constants one should consider. The $RG$ equations are derived in Sec.~\ref{RGE} and the calculation of their coefficients   in the two-loop approximation of the $\phi^4$ theory is detailed in Sec.~\ref{secRGE}. The discussion of our results as well as our conclusions are presented in Sec.~\ref{conc}.

\section{General Considerations} \label{general}
\subsection{Relevant scales}\label{relevant}
As mentioned above, it is essential in a first step to recall  the relevant scales which appear in the calculation of any 
 amplitude, and to specify their interpretation.  
These scales should be organized in two  different sets~:\\

\noindent{\it i) Physical conditions}\\
We can identify two different scales~:
\begin{itemize}
\item The energy/momentum scales which specify the kinematical conditions of the experiment.  We should consider in principle a whole set of  scales in order to completely fix the kinematics of the initial and/or final states in inclusive, semi-inclusive or exclusive processes. For simplicity we denote them by $Q$.
\item The  masses of all the particles associated with the elementary  degrees of freedom  present in the Lagrangian. For simplicity we denote them by $m$. We assume in this study that $m$ is strictly nonzero. 
\end{itemize}
{\it ii) Arbitrary scales}\\
Two scales are  a priori arbitrary~:
\begin{itemize}
\item The scale associated to the regularization procedure. We call this scale the {\it regularization scale} and denote it symbolically by $\eta$.  It is characteristic of the regularization method which is used to give a mathematical well defined meaning to the local bare Lagrangian we start from. As we shall argue in this study, this scale should be considered as {\it dimensionless} when using $DR$. This is already the case using $TLRS$ \cite{GW}.
\item The energy/momentum scale at which an experiment should be performed in order to fix the value of the parameters of the Lagrangian. Like $Q$, it is more precisely a set of scales, but for simplicity we shall denote them by $M$. We  call this scale the {\it renormalization point} since it fixes the kinematical point where the physical renormalization of the bare parameters is calculated. 
\end{itemize}
Note that the regularization scale $\eta$ is defined prior to the choice of a renormalization scheme, while the renormalization point $M$ is chosen independently of the regularization procedure and renormalization scheme which are used \footnote{ The term ``renormalization'' in ``renormalization point''  refers here to the physical renormalization of the bare parameters, as it appears for instance in any interacting many-body system, and do not refer whatsoever to the choice of a particular ``renormalization scheme''.}.  A change of renormalization scheme can, to some extent, be analyzed in terms of a {\it re-definition} of the regularization scale $\eta$. But this does not mean that $\eta$ is intrinsically linked to a particular renormalization scheme: $\eta$ should be first {\it defined} independently of any renormalization scheme before it can be trivially {\it redefined} to account for  a different  scheme. This redefinition just shows the equivalence between these renormalization schemes.

We would like to emphasize  the quite different nature of these two arbitrary scales. The first one, $\eta$, and the $RG$ equation associated to it denoted by $RGE(\eta)$, is just required by the internal consistency of the theoretical framework which is used, independently of the relevance of the Lagrangian we start from to describe the dynamics of the system under consideration.  The invariance condition associated to it is  completely independent of the real physical dynamics of this system. It is only related to the intrinsic property of a theoretically calculated physical observable - this one should be independent of the regularization scale -  but by no means to the ability of this theoretical quantity to accurately describe an experimental physical measurement in a large domain of $Q$, {\it i.e.} to describe real physical dynamics. We shall call this invariance a {\it weak invariance condition. One should not  give it any physical significance}. The dimensionless nature of this scale prevents indeed any attempt to {\it interpret} it in terms of a particular physical scale like $Q$.

Since the scale $M$ refers to an experimental measurement, the independence of physical observables on $M$, leading to its $RG$ equation $RGE(M)$, on the other hand, is {\it true and only true} if the dynamics of the system under consideration is reasonably well accounted for by the Lagrangian we start from, in a given domain of energy and provided the calculations are performed with an appropriate accuracy. 
{\it This $RG$ equation is therefore the only one which should be used if true physical patterns are searched for}. It is independent of the choice of a particular regularization procedure or  a particular renormalization scheme. In this sense it is universal, in contrast to $RGE(\eta)$. The choice of M is a question of taste and feasibility of the experiment.
This invariance condition is called a {\it strong invariance condition} since it requires to reproduce  experimental physical measurements in a large domain of $Q$, {\it i.e.} to reproduce physical reality within some accuracy  in a given energy range. {\it Note that the strong invariance condition implies the weak one, but the reverse is of course not true}. This confirms the  different nature of $\eta$ and $M$.

\subsection{Bare and physical coupling constants} \label{bare}
\begin{figure}[bt] 
\centerline{\includegraphics[width=15pc]{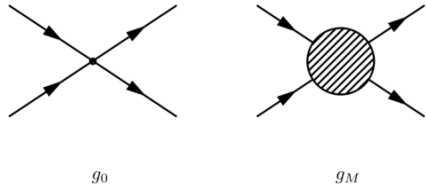}} 
\begin{center}
\caption{\label{fig1} \small Bare ($g_0$) and physical ($g_M$) coupling constants in a $\phi^4$-type theory.  }
\end{center}
\end{figure}

We can identify {\it a priori} two fundamental and uniquely defined coupling constants. The first one is the bare coupling constant\footnote{We restrict ourself for simplicity in this study to only one interaction term in the bare Lagrangian. The extension to many couplings can be done similarly.}, denoted by $g_0$, as it appears in the bare Lagrangian we start from. The second one is the physical coupling constant calculated at the scale $M$ and denoted by $g_M$. It can, in principle, be measured experimentally, except for well known  cases as we shall comment at the end of this section. These two coupling constants are represented schematically by the diagrams of Fig.~\ref{fig1}, in the ideal case of a 4-legs coupling constant. 
The physical coupling constant is  determined directly from a measurement of the scattering amplitude at the two particular kinematical scales $s_0$ and $t_0$, with  $M\equiv(\sqrt{s_0},\sqrt{\vert t_0 \vert})$. 
Note that these two coupling constants are well identified prior to the choice of any  regularization procedure or any  renormalization scheme. 

By definition, $g_M$ does depend on $M$ only. Since it is a physical observable, it cannot depend on the arbitrary regularization scale $\eta$. We shall thus denote it by $g_M(M)$. On the contrary, $g_0$ should depend {\it a priori} on $\eta$ but not on $M$. We denote it by $g_0(\eta)$. It is clear that $g_0$ cannot depend on $M$, otherwise any physical observable, calculated at an arbitrary value of $Q$, would depend on $M$, since any amplitude, amputated from the bare coupling constant, does not depend explicitly on $M$ at any order of perturbation theory.
Its dependence on $\eta$ originates from the fact that the Lagrangian {\it density}  - which is not a physical observable - should depend on the way it is (mathematically well) defined, {\it i.e.} on the regularization procedure.  Using TLRS for instance \cite{GW},  it should depend on the way the fields are defined in order to give a mathematical meaning to the  product of field operators - considered as operator-valued distributions - at the same point. 
This exhibits naturally a dimensionless  parameter $\eta$, associated with the scaling properties in the ultraviolet  limit. Using $DR$, it should  depend on the dimensionality of the space-time, and hence on the ``unit of mass'', called $\mu$, necessary to give the right dimensionality to the coupling constant  in $D$ dimensions \cite{thooft}.   This ``unit of mass'' should be defined with respect to a single measured mass scale, which we call $M_0$. We thus write it as 
\begin{equation} \label{unit}
\mu\equiv \eta\ M_0, 
\end{equation}
where $\eta$ is an arbitrary {\it dimensionless} scale. Since $M_0$ is a measured quantity, it can always be chosen to be non zero. Note that one should not assume {\it a priori} any given dependence (or independence) of $g_0$ on the regularization scale $\eta$, but just calculate it  from first principles, as we shall show in Sec.~\ref{secRGE}. 
By many aspects, $M_0$ should be considered as a true unit of mass, by analogy with the units of length and  time. Any measure of a nonzero mass or kinematical variable, like for instance $m$ or $M$, implicitly assumes the choice of a given unit of mass $M_0$. Since $\mu$ is not a measured quantity, $M_0$ should be specified by hand.  

We shall illustrate our purpose in the following by using $DR$ with two different renormalization schemes. As it is well known, one considers in this case an auxiliary coupling constant, called  also ``renormalized'' and denoted by $g_R$. This coupling constant is defined by adding  appropriate counterterms to the original Lagrangian in order to cancel, in perturbation theory, the poles in $\epsilon^n$, up to an arbitrary constant which may depend a priori on $\eta$ and/or $M$. In the minimal subtraction ($MS$) scheme, and its variants, one just removes by hand these poles, up to irrelevant constants. In the on-mass shell ($OMS$) scheme, the renormalized coupling constant is identified  with the physical one $g_M$, assuming perturbation theory is adequate in order to calculate $g_M(M)$. Note  that $g_0$ and $g_M$ can be defined or measured for any value of these coupling constants, in the perturbative as well as non-perturbative regimes, while $g_R$ can only be defined in the perturbative domain through the construction of the counterterms.

The two coupling constants $g_0(\eta)$ (or equivalently $g_R(\eta)$ in the $MS$ scheme) and $g_M(M)$  refer to two separate properties. The first one is a property of the  bare local Lagrangian we start from, without any knowledge of the physical state which is realized in Nature in a given domain of energy. This physical state is a solution of the equations of motion. In contrast, the second one  is, by definition, a property of this physical state at the energy scale $M$. This distinction will be of utmost importance for systems showing spontaneous symmetry breaking, like $QCD$, as we shall comment in Sec.~\ref{conc}.

The above discussion should, of course, be
 applied with some care for $QCD$ since the relevant degrees of freedom are not asymptotic states in this case. One cannot, therefore, consider a direct measurement of $g_M$. It is however well known \cite{grun,deur} that one can consider alternatively other simple observables, like the $e^+ e^-$ annihilation, the DIS cross-sections, or the Bjorken sum rule, in order to define an effective charge, up to scale-independent factors.  This is also the case when the coupling constant cannot be determined experimentally at an arbitrary value for $M$ with on-shell external particles, like in $QED$ for instance \cite{gell}. All the derivations we detail in this study apply as well in these cases. A similar strategy is of course also possible for the mass parameter.

\section{The Renormalization Group Equations} \label{RGE}
\subsection{Derivation of the equations}
The $RG$ equations $RGE(\eta)$ and $RGE(M)$ are derived from the invariance of the $S$ matrix with respect to the variation of the arbitrary scales $\eta$ and $M$, respectively. 
Doing this, one should first specify which parameters should be kept fixed.  According to our discussion in Subsec.~\ref{bare}, $RGE(\eta)$ should be derived with the physical parameters kept fixed, since these are independent of $\eta$, while $RGE(M)$ should be derived with the bare parameters kept fixed, since these are independent of $M$.
Note that the invariance of the $S$ matrix is only true in the physical world, {\it i.e.} at $\epsilon=0$. We thus get, for the $\eta$ dependence,
\begin{equation} \label{RGEeta}
\left[ \eta \frac{\partial}{\partial \eta}+\beta_\eta^R\frac{\partial}{\partial g_R}+m_R\ \gamma_{m,\eta}^R\ \frac{\partial}{\partial m_R}\right] S=0,
\end{equation}
for a generic renormalization scheme $R$, with
\begin{equation}
\beta_\eta^R=\eta\left. \frac{\partial g_R}{\partial \eta}\right \vert_{g_M,m_M}, \ \ \ \gamma_{m,\eta}^R=\frac{\eta}{m_R}\left. \frac{\partial m_R}{\partial \eta}\right \vert_{g_M,m_M},
\end{equation}
where $m_M$ is defined below.
This is the $RG$ equation for $\eta$, denoted by $RGE(\eta)$. It is well defined at $\epsilon=0$ since, by construction, the renormalized parameters $g_R$ and $m_R$ are finite.
Note that it is not possible to derive a similar equation involving the renormalized $n$-point Green's function $G_R^{(n)}$, written as 
\begin{equation} \label{green}
G_R^{(n)}=Z_\phi^{-n/2}\ G_0^{(n)},
\end{equation}
where $Z_\phi$ is the usual field renormalization factor. The bare parameters $g_0$ and $m_0$ do indeed depend {\it a priori} on the regularization scale $\eta$ and $G_0^{(n)}$ is not a physical observable. 

The $RGE(M)$ equation is derived similarly, with
\begin{equation} \label{goms}
\left[ M \frac{\partial}{\partial M}+\beta_M\frac{\partial}{\partial g_M}+m_M\ \gamma_{m,M}\ \frac{\partial}{\partial m_M} \right] S=0,
\end{equation}
and
\begin{equation} \label{Manomalous}
\beta_M=M\left. \frac{\partial g_M}{\partial M}\right \vert_{g_0,m_0},\  \gamma_{m,M}=\frac{M}{m_M}\left. \frac{\partial m_M}{\partial M}\right \vert_{g_0,m_0}. 
\end{equation}
An identical equation for the $n$-point Green's function $G_{OMS}^{(n)}$ can be also derived in this case since in the $OMS$ scheme the $S$ matrix is equivalent to the Green's function.

The $RG$ equation (\ref{RGEeta}) involves the $S$ matrix only, with no reference whatsoever on the anomalous dimension. It concerns for instance the renormalized coupling constant $g_{MS}(\eta)$ and mass $m_{MS}(\eta)$ in $DR$ with the $MS$ renormalization scheme (denoted by $DR+MS$ in the following). The $RG$ equation (\ref{goms}) refers to the $S$ matrix or to the $n$-point Green's function in the $OMS$ scheme using $DR$. It concerns the physical coupling constant $g_M(M)$ and pole mass $m_M$.
Note that when dealing with particles which are asymptotic states, the pole mass $m_M$ is by definition a physical parameter, with $m_M\equiv m$. It is thus independent of the renormalization point $M$, so that $\gamma_{m,M}\equiv 0$ in this case.

\subsection{Calculation of the coefficients}
\subsubsection{The $\beta$ functions}
We shall illustrate our purpose in the following using the $\phi^4$ theory. The  physical coupling constant $g_M$ - calculated at the two kinematical variables $(s_0,t_0)$ necessary to completely fix the conditions of the reaction - is defined from the physical scattering amplitude according to
\begin{equation}\label{T0}
T(s_0,t_0) \equiv  g_M(M).
\end{equation}
All  physical observables correspond, by definition, to $\epsilon=0$. This is  the case for $g_M$ in Eq.~(\ref{T0}). It is thus dimensionless. To avoid any confusion, we shall denote by a hat all dimensionful coupling constants defined in $DR$ at $\epsilon \neq 0$. Only the  dimensionful bare coupling constant $\hat g_0$ is in fact relevant for the following discussion.

In terms of the coupling constants $\hat g_0$ and $g_R$, we have
\begin{subequations} \label{alr}
\begin{eqnarray}
T(s_0,t_0) &=&\lim_{\epsilon \to 0} \left[\hat g_0  \left[1+\Lambda_0(M,\eta)\right] Z_0^2\right]   \\
&=&g_R \left[1+\Lambda_R(M,\eta)\right] Z_R^2\label{alra}, 
\end{eqnarray}
\end{subequations}
where the vertex functions $\Lambda$ are decomposed according to  
\begin{subequations} \label{PiQ2}
\begin{eqnarray}
\Lambda_0&=&\Delta \Lambda_\epsilon+ \Lambda_{F}^0, \label{PiQ2a}\\
\Lambda_R&=&\Delta \Lambda_\epsilon+ \Delta \Lambda_{CT}^R + \Lambda_{F}^R.\label{PiQ2b}
\end{eqnarray}
\end{subequations}
The usual field strength renormalization constants are denoted by $Z$. In these equations, the vertex function $\Lambda_0$ and field strength $Z_0$ depend on $g_0$ and $m_0$, while $\Lambda_R$ and $Z_R$ depend on $g_R$ and $m_R$. They both depend explicitly on $\eta$ and $M$.  The term $\Delta \Lambda_\epsilon$ in the above expressions refers to the divergent piece of the vertex functions - apart from irrelevant constants independent of both $\eta$ and $M$ - while $\Delta \Lambda_{CT}^R$ stands for the contribution from the corresponding counterterms. The $\Lambda_F$ terms are the leftover finite pieces. The  constants $Z_0$ and $Z_R$ are defined similarly.  Since $g_R$, $\Lambda_R$ and $Z_R$ are by construction finite, Eq.~(\ref{alra}) is well defined at $\epsilon = 0$. Note that the $\Lambda_F$ terms have {\it a priori}  two different  contributions: the first one is the leftover finite piece of an otherwise divergent contribution, the second one is a purely convergent contribution. While the first one does depend explicitly on the regularization scale $\eta$, the second one does not. It depends however explicitly on $M$ and contribute therefore to $RGE(M)$ but not to $RGE(\eta)$.

Using Eqs.~(\ref{T0},\ref{alr}), the various coupling constants are related, at $\epsilon = 0$, by
\begin{subequations} \label{gm0} 
\begin{eqnarray} 
g_M(M)&=&\lim_{\epsilon \to 0} \left[ \hat g_0(\eta) \bar Z_0(M,\eta)\right] \label{gm0a} \\
&=&g_{R}(\eta)\bar Z_{R}(M,\eta),\label{gm0b} 
\end{eqnarray}
\end{subequations}
with, in a schematic notation,
\begin{equation}
\bar Z \equiv  \left(1+ \Lambda\right) Z^2.
\end{equation}
In the $DR+MS$ scheme, we have simply
\begin{equation}\label{CTi}
\Delta \Lambda_{CT}^{MS}=-\Delta \Lambda_\epsilon.
\end{equation}
The $\beta_\eta^{R}$ function can be directly calculated starting from the finite contribution to the vertex functions. This is at variance with the common derivation where $\beta_\eta^{R}$ is calculated from the divergent contribution to the amplitude, once the hypothesis that the bare coupling $\hat g_0$ be invariant with respect to $\eta$ is made. As argued in the introduction, the behavior of $\hat g_0$ as a function of $\eta$ should not be fixed {\it a priori }but calculated starting from physical requirements. 

From (\ref{gm0b}), and the invariance of $g_M$ with respect to $\eta$, we get
\begin{equation} \label{gMS}
\beta_{\eta}^R= -\frac{g_R}{\left[\bar Z_R+g_R \frac{\partial \bar Z_R}{\partial g_R}\right]} \left[ \eta \frac{\partial \bar Z_R}{\partial \eta}+m_R\gamma_{m,\eta}^R\frac{\partial \bar Z_R}{\partial m_R}\right].
\end{equation}
The calculation of $\beta_M$  follows from a similar derivation. 
From  the invariance of $g_R(\eta)$ with respect to $M$, we have simply, always at $\epsilon=0$,
\begin{equation} \label{alphaM}
\beta_M= g_R \ M\frac{\partial \bar Z_R}{\partial M}.
\end{equation} 
The derivative with respect to $M$ in the above equation is a shorthand notation for the derivative with respect to $\sqrt{s_0}$ or $\sqrt{\vert t_0 \vert }$, separately, in the $\phi^4$ theory we are considering in this study. 

\subsubsection{The $\gamma_m$ function}
The calculation of the $\gamma_m$ coefficient follows  from the same spirit.  By definition, the physical observable  we should refer to is the physical (pole) mass $m$, defined, in the case of a scalar boson, by
\begin{equation} \label{mo}
m^2=\lim_{\epsilon \to 0} [m_0^2+\Sigma_0(p^2=m^2,\eta)],
\end{equation}
where $m_0$ is the bare mass of the boson and $\Sigma_0$ the self-energy correction.  We decompose $\Sigma_0$ as
\begin{equation}
\Sigma_0= \Delta \Sigma_\epsilon+ \Sigma_{F}^0.
\end{equation}
The bare mass does depend {\it a priori} on the arbitrary regularization scale $\eta$, like the bare coupling constant $\hat g_0$ does, as argued in Subsec.~\ref{bare}. It does not depend on $M$. The physical mass does  depend, by definition, on neither $\eta$ nor $M$. In an arbitrary renormalization scheme $R$, the renormalized mass is defined by
\begin{equation} \label{mr}
m^2=m_R^2+\Sigma_R(p^2=m^2,\eta),
\end{equation}
with
\begin{equation}\label{sigR}
\Sigma_R=\Delta \Sigma_\epsilon+\Delta \Sigma_{CT}^R+\Sigma_{F}^R.
\end{equation}
The contribution from the counterterms, denoted by $\Delta \Sigma_{CT}^R$, may depend {\it a priori}  on  $\eta$ and/or $M$, like $\Delta \Lambda_{CT}^R$ in (\ref{PiQ2}b).
For the $DR+MS$ scheme, we have, similarly to (\ref{CTi}),
\begin{equation}\label{sigCT}
\Delta \Sigma_{CT}^{MS}=-\Delta \Sigma_\epsilon.
\end{equation}
We thus get immediately, from (\ref{mr}),
\begin{equation} \label{gmr}
\gamma_{m,\eta}^R=-\frac{1}{2m_R^2\left[1+\frac{1}{2m_R}\frac{\partialÊ\Sigma_R}{\partial m_R}\right]} \left[\eta\frac{\partial \Sigma_R}{\partial \eta}+\beta_\eta^R\frac{\partial\Sigma_R}{\partial g_{R}}\right].
\end{equation}
%

\section{Simple example} \label{secRGE}
In the following, the index $R$ refers always to the $DR+MS$ scheme, for simplicity of the notations.

\subsection{$\phi^4$ theory in the two-loop approximation} \label{phi4}
The behavior of the renormalized parameters $g_R$ and $m_R$  can be directly obtained at $\epsilon=0$ using (\ref{gMS}) and (\ref{gmr}). From the calculations of Ref.~\cite{klei}, we  get for the $\beta$ and $\gamma_m$ coefficients, at the one-loop order
\begin{subequations}
\begin{eqnarray}
\beta_\eta^{(1)}&=&3\frac{g_R^2}{(4\pi)^2}, \label{br1}\\
\gamma_{m,\eta}^{(1)}&=&\frac{1}{2}\frac{g_R}{(4\pi)^2},
\end{eqnarray}
\end{subequations}
and at the two-loop order
\begin{subequations}
\begin{eqnarray}
\beta_\eta^{(2)}&=&-\frac{17}{3}\frac{g_R^3}{(4\pi)^4},\\
\gamma_{m,\eta}^{(2)}&=&-\frac{5}{12}\frac{g_R^2}{(4\pi)^4}.
\end{eqnarray}
\end{subequations}
The full scale-dependent contributions to $g_R$ and $m_R$ are given explicitly below, in Eqs.~(\ref{deltag}), (\ref{deltag2}) and (\ref{deltam}).
While the physical hypothesis we assume in this study to calculate the $\beta_\eta$ and $\gamma_{m,\eta}$ functions  - starting from the invariance properties of physical observables - are quite different from the one used up to know - assuming the independence of the bare parameters as a function of the regularization scale - we get the same, scale-independent, $RG$ coefficients to this non-trivial order. 

Let us analyze in more details the connection between these two hypothesis, called hypothesis $1$ and $2$, respectively.
In the standard derivation (hypothesis $2$), $\hat g_0$ as well as $m_0$  are assumed to be independent of $\mu$. Since $\hat g_0$ is dimensionful, it should thus be written in this case as
\begin{equation} \label{hatg0b}
\hat g_0=g_0^{st}\ (M_0)^\epsilon,
\end{equation}
where $M_0$ is a given unit of mass, independent of $\mu$, as defined for instance in (\ref{unit}). Since in this hypothesis $\hat g_0$ is assumed to be independent of $\mu$, $g_0^{st}$ is also independent of $\mu$. With this hypothesis, the calculation of any physical observable, and hence also of $g_M(M)$, is thus trivially independent of $\mu$: the hypothesis $2$ is compatible with the hypothesis $1$. 

This, of course, does not imply that both hypothesis  are equivalent! The invariance of physical observables in hypothesis $1$ - calculated by definition at $\epsilon=0$ - cannot induce any constraint on the behavior of $\hat g_0$ at $\epsilon \neq 0$ since $\hat g_0$ is singular in the limit $\epsilon \to 0$ in $DR$. On the other hand, a given hypothesis on the behavior of $\hat g_0$ at $\epsilon \neq 0$ - as done for instance in the standard approach in hypothesis $2$ - can induce a constraint on the physical coupling constant $g_M$ at $\epsilon = 0$ since any theoretical physical observable should be finite in the limit $\epsilon \to 0$. 

Let us see  how this works in practice in a simple one-loop calculation. The bare coupling constant is related to the renormalized one by the standard relation
\begin{equation} \label{hatg0}
\hat g_0=\frac{Z_1}{Z_2^2}\ g_R\ \mu^\epsilon,
\end{equation}
where $Z_1$ and $Z_2$ are the usual vertex and field renormalization constants. This relation is only defined at $\epsilon \neq 0$ since $\hat g_0$ as well as $Z_{1}$ and $Z_2$ have poles in $\epsilon$, in $DR$. Our derivation starting from (\ref{gMS}) leads to the value (\ref{br1}) of the $\beta_\eta$ function calculated at $\epsilon = 0$. On the other hand, starting from (\ref{hatg0}) and with the hypothesis $2$ that $\hat g_0$ is independent of $\mu$, we get, at $\epsilon \neq 0$, the standard expression for the $\beta_\eta$ function, denoted by $ \beta_\eta^{st}$, 
\begin{equation} \label{standg}
\beta_\eta^{st}=-\epsilon  g_R + 3\frac{g_R^2}{(4\pi)^2}.
\end{equation}
The term $-\epsilon g_R$ in (\ref{standg}) is essential in order to get the independence of $\hat g_0$ as a function of $\eta$, from Eq.~(\ref{hatg0}). 

How could we recover Eq.~(\ref{standg}) from our analysis? This requires to extend the physical coupling constant in (\ref{gm0a}) at $\epsilon \neq 0$. {\it This extension is not unique}. It can be done, for instance, in two different ways, according to
\begin{subequations}
\begin{eqnarray}
(a)\ \ &g_M(M)& \to g_M(M)\ \mu^\epsilon,\label{extb}\\
(b)\ \ &g_M(M)& \to g_M(M)\ (M_0)^\epsilon.\label{exta}
\end{eqnarray}
\end{subequations}
The case $(a)$ is the analogue of the definition of the dimensionless renormalized coupling constant in $DR$, with $\hat g_R=g_R\ \mu^\epsilon$, while  case $(b)$ is the analogue of the extension (\ref{hatg0b}) used in the standard derivation. From (\ref{gm0a}) extended at $\epsilon \neq 0$, we can thus calculate the behavior of $\hat g_0$, characterized by 
\begin{equation}
\hat \beta_0=\eta \frac{\partial \hat g_0}{\partial \eta},
\end{equation}
from the invariance of $g_M$ with respect to $\eta$. We get immediately, with  these two possible extensions at $\epsilon \neq 0$, 
\begin{subequations}
\begin{eqnarray}
\beta_\eta^{(1,a)}&=&3\frac{g_R^2}{(4\pi)^2}\label{br1b},\\
\beta_\eta^{(1,b)}&=&-\epsilon g_R+3\frac{g_R^2}{(4\pi)^2}.
\end{eqnarray}
\end{subequations}
Note that with our choice $(a)$, the $\beta_\eta^{(1,a)}$ coefficient has no contribution $-\epsilon g_R$, eventhough it is calculated at $\epsilon \neq 0$.
With these values, we can then calculate directly $\hat \beta_0$ from (\ref{hatg0}) without any new hypothesis. We get :
\begin{subequations}
\begin{eqnarray}
\hat \beta_0^{(1,a)}&=&\mu^\epsilon \left[\epsilon g_R+6\frac{g_R^2}{(4\pi)^2} \right], \label{hatb0b} \\
\hat \beta_0^{(1,b)}&=&0.\\
\end{eqnarray}
\end{subequations}
While we recover our results (hypothesis $1$) with  choice $(a)$, leading to a nonzero contribution for $\hat \beta_0$, we recover with  choice $(b)$ the standard hypothesis (hypothesis $2$). Defining the bare dimensionless coupling constant $g_0$ by the natural expression
\begin{equation} \label{defg0}
\hat g_0=g_0\ \mu^\epsilon,
\end{equation}
we get, always at $\epsilon \neq 0$,
\begin{equation}\label{b01b}
\beta_0^{(1,a)}\equiv \eta \frac{\partial g_0}{\partial \eta}=3\frac{g_R^2}{(4\pi)^2}.
\end{equation}
It is equal to $\beta_R^{(1,a)}$ which is also equal to the value of $\beta_R$ calculated in (\ref{br1}) at $\epsilon=0$. We recover  with  choice (a) the universality of the $\beta$ function in the one-loop approximation, including the bare coupling constant.

\subsection{Decoupling properties} \label{decoup}
In order to exhibit the connection between the $M$-dependence of $g_M$ and the $\eta$-dependence of $g_R$, let us write Eq.~(\ref{gm0b}) in a more transparent way,
\begin{equation} \label{dec}
g_M(M)\left[1+\Delta g_M(g_M,M)\right]=g_R (\eta)\left[1 + \Delta g_R(g_R,\eta)\right] \equiv \bar g.
\end{equation}
We have expressed in this equation all $\eta$-dependent contributions in terms of $g_R$, while the $M$-dependent contributions are expressed in terms of $g_M$, as well as the constant terms. This separation is a necessary condition in order to guarantee that the renormalized and physical coupling constants calculated from Eq.~(\ref{gm0b}) depend only on $\eta$ and $M$, respectively. A similar decoupling occurs also when the renormalized coupling constant, in the $DR+MS$ scheme, is replaced by the bare one defined in (\ref{defg0}),  since the relation between the two coupling constants is M-independent, according to (\ref{hatg0}). The decoupling equation (\ref{dec}) exhibits also an invariant parameter which we called $\bar g$. It is independent of both $M$ and $\eta$, but depends implicitly on the choice of the unit of mass $M_0$. 

This decoupling of the $\eta$- and $M$-dependences is of course not trivial. We can easily see  that it is indeed the case in a simple one-loop calculation, with, always in the $\phi^4$ theory \footnote{In all these calculations, we have taken $M_0\equiv m$. We also included irrelevant constants in a redefinition of $\eta$, like in the $\overline {MS}$ scheme.},
\begin{subequations} \label{deltag}
\begin{eqnarray}
\Delta  g_M^{(1)}(g_M,M)&=& \frac{1}{2} \frac{g_M}{(4\pi)^2}I(s_0,t_0),\\
\Delta  g^{(1)}_R(g_R,\eta)&=&-\frac{3}{2} \frac{g_R}{(4\pi)^2}\mbox{Log}\ (\eta^2),
\end{eqnarray}
\end{subequations}
with
\begin{equation}
I(s_0,t_0)=J(s_0)+J(t_0)+J(4m^2-s_0-t_0),
\end{equation}
and
\begin{equation}
J(s)=\int_0^1 \mbox{Log} \left[ \frac{m^2}{m^2-s x(1-x)}\right] dx.
\end{equation}
In the two-loop calculation within the $\phi^4$ theory we have done in this study, we checked that this is also the case, with, for the simplest and most interesting expression $\Delta g^{(2)}_R$,
\begin{equation} \label{deltag2}
\Delta g^{(2)}_R(g_R,\eta)= \frac{1}{12} \frac{g_R^2}{(4\pi)^4}\left[ 34\ \mbox{Log}\ (\eta^2)
+27 \left[\mbox{Log}\ (\eta^2)\right]^2\right].
\end{equation}
A similar separation occurs also for the renormalized mass. We get in this case
\begin{equation} \label{deltam}
m^2=m_R^2\left(1-\frac{1}{2}\frac{g_R}{(4\pi)^2}\left[\mbox{Log}(\eta^2)+1\right]
+\frac{1}{12}\frac{g_R^2}{(4\pi)^4}\left[17\ \mbox{Log}(\eta^2)+ 6\left[\mbox{Log}(\eta^2)\right]^2\right]\right),
\end{equation}
up to terms independent of $\eta$ at order $g_R^2$, which are irrelevant for the calculation of $\gamma_m$ at that order.
This expansion is reminiscent, in a one-loop calculation, to what has been found using $TLRS$ in Ref.~\cite{FT} for the calculation of radiative corrections to the Higgs mass in the Standard Model.

Note that this decoupling is also true when the perturbative series, calculated in the one-loop approximation, is resummed to all orders, like in $QED$ for instance. We get in this case, with $M^2\equiv Q^2$,
\begin{equation}
\alpha_M(M)=\frac{\alpha_R(\eta)}{1+\alpha_R(\eta)\left( A_M+A_\eta\right)},
\end{equation}
and
\begin{eqnarray}\nonumber
A_M&=&-\frac{2}{\pi} \int_0^1 x(1-x)\ \mbox{Log}\left[\frac{m^2+x(1-x)M^2}{m^2}\right]dx,\\
 A_\eta&=&\frac{1}{3\pi}\ \mbox{Log}(\eta^2).\nonumber
\end{eqnarray}
We thus have the following decoupling relation:
\begin{equation}
\frac{\alpha_M(M)}{1-A_M\  \alpha_M(M)}=\frac{\alpha_R(\eta)}{1+A_\eta\  \alpha_R(\eta)}\equiv \bar \alpha,
\end{equation}
with
\begin{equation}
\bar \alpha = \alpha_M(M=0)=\alpha_\eta(\eta=1)\equiv \alpha
\end{equation}
where $\alpha$ is the fine structure constant.

\section{Discussion and conclusions} \label{conc} 
Starting from a well-behaved local Lagrangian regularized by $DR$, we have separated  in this study two different $RG$ equations. 
The first one, called $RGE(\eta)$,  concerns the $S$ matrix only. It is based on the invariance of any theoretically calculated physical observable on the  regularization scale $\eta$. 
The second one, called $RGE(M)$, concerns also the $S$ matrix. It is based on the invariance of any  physical observable on the  renormalization point $M$ used to fix, from an experimental measurement, the value of the bare or renormalized parameters of the Lagrangian. We recall that this latter invariance is true and only true if the Lagrangian we start from is adequate to describe the physical processes under consideration in a given domain of energy and with a given accuracy ({\it strong invariance condition}).
These two $RG$ equations are complementary: $RGE(M)$ entails the full physical significance of the dynamics of the system, while $RGE(\eta)$ guaranties the theoretical consistency of the calculation, independently of the relevance of the Lagrangian we start from to describe the full dynamics of the system under consideration ({\it weak invariance condition}). 
While a strong invariance condition implies a weak one, the reverse is of course not true. Note also that $\eta$ is a single scale while $M$ can be a multiple scale, like  in the $\phi^4$ theory.

In view of our results, it is interesting to come back to the usual hypothesis made in the literature for the calculation of the coefficients of the $RG$ equations. It is assumed in this case  that the bare parameters - the bare dimensionful coupling constant $\hat g_0$ and the bare mass $m_0$ - are independent of the dimensionful scale $\mu$ of $DR$. Contrarily to the derivation of the $RG$ equation itself -  starting for instance from the invariance of the $S$ matrix - this hypothesis does not rely on any requirement based on physical principles. As mentioned in Subsec.~\ref{bare}, the bare parameters are not physical observables, and should thus depend {\it a priori}  on the way the local Lagrangian we start from is regularized in order to be well-behaved from the very beginning.

The relationship between the two approaches can be understood as follows. Any hypothesis on $\hat g_0$ cannot rely on physical grounds since it necessitates to extend the coupling constant, either physical or renormalized, at $\epsilon \neq 0$, as it can be seen, for instance, from Eqs.~(\ref{gm0}). This extension is not unique. The standard approach assumes implicitly an extension of the physical coupling constant given by (\ref{exta}), {\it i.e.} independent of $\mu$, by analogy to the construction of $\hat g_0$ in (\ref{hatg0b}). On the other hand, an extension of $g_M$ at $\epsilon\neq 0$ analogous to the construction of the renormalized coupling constant in $DR$, with (\ref{extb}), leads to a scale-dependent bare coupling constant $\hat g_0$, with a non zero $\beta$ function $\hat \beta_0$, given by (\ref{hatb0b}). In this case, the (dimensionless) bare coupling constant and the renormalized one have the same regularization scale dependence in a simple one-loop calculation in the $\phi^4$ theory, as shown in Eqs.~(\ref{br1b}) and (\ref{b01b}). Moreover, all relevant coupling constants $g_0, g_R$ and $g_M$ are extended in this case, at $\epsilon \neq 0$, by the universal relation
\begin{equation}
\hat g = g \ \mu^\epsilon.
\end{equation}
This is indeed a natural condition since the extension of the Lagrangian at $\epsilon \neq 0$ should be done independently of the way this Lagrangian is defined in terms of the bare, renormalized or physical fields. With this universal extension, the dimensionful bare coupling constant $\hat g_0$, like the bare mass $m_0$, does depend on the regularization scale, contrarily to what is usually assumed in the literature.

We argued in this study that the relevant regularization scale in $DR$ should be dimensionless. While this is definitely not ruled out by any physical or mathematical principle, the dependence of $g_{R}$ on a dimensionless variable is  indeed a necessary condition in order to get a non zero massless limit for its $\beta$ function. Since the $\beta_\eta^{R}$ function is mass independent, it should  be non zero in the massless limit. The renormalized coupling constant $g_{R}(\eta)$ should thus be  scale-dependent in this limit. This is however only possible if the regularization scale $\eta$ is a dimensionless quantity. This would not be the case  with a dimensionful regularization scale parameter $\mu$ since in the absence of any other parameter with the dimension of a mass at the level of the Lagrangian, the dimensionless quantity $g_{R}$ should be independent of $\mu$, with hence a zero $\beta$ function.

The behavior of $g_M(M)$ in the massless limit, on the other hand, proceeds differently since $\eta$ and $M$ are of different nature, as already argued in Subsec.~\ref{relevant}. The regularization scale is completely arbitrary and nonphysical, and it should not be fixed in any way by a physical measurement. Once a unit of mass is defined by a single measurement called $M_0$ in (\ref{unit}), this arbitrariness can only be accounted for by a dimensionless quantity which we have called $\eta$. On the other hand, $M$ is also arbitrary, but it is fixed, by definition, by a physical measurement. This measurement implicitly assumes a given unit of mass, so that measuring $M$ or $m$ fixes also equivalently the value of $M_0$. 

It is interesting, in this context, to discuss the case of massless $QCD$. On the one hand, the dimensionless renormalized coupling constant $g_{R}(\eta)$ in the $DR+MS$ scheme does depend on the dimensionless regularization scale $\eta$.  It cannot therefore depend  on the $QCD$ parameter $\Lambda_{QCD}$ since there is no other parameter with the dimension of a mass to compensate at the level of the renormalized Lagrangian. This is moreover consistent with the fact that this Lagrangian indeed knows nothing about the spontaneous breaking of chiral symmetry  and the appearance of condensates in physical systems, in a given domain of energy. On the other hand, the effective coupling constant $g_M(M)$, extracted for instance from the Bjorken sum rule \cite{deur}, does depend on the mass parameter $\Lambda_{QCD}$ which can be associated to the spontaneous breaking of chiral symmetry, with a value of the order of the mass scale of the quark condensate, for instance. This effective coupling constant does depend on $M$ through the dimensionless ratio $M/\Lambda_{QCD}$, showing asymptotic freedom for large $M$. Note that in absence of any condensates, $g_M$ should be independent of $M$ since there would be no mass parameter to compensate for $M$. 

The value of $\Lambda_{QCD}$ extracted from the analysis of $g_M$  is thus universal in the sense that it does  depend neither on the regularization procedure nor on the renormalization scheme  which have been chosen in order to perform the calculations, since a physical observable should be independent of them. It does not depend either on any mass threshold since $g_M(M)$ incorporates {\it de facto} the right decoupling properties, as we shall comment below. In the one-loop approximation, it does not depend also on the process used to extract the effective coupling constant. It should therefore be determined in the high energy limit in order to minimize this model dependence at higher orders, thanks to asymptotic freedom \cite{deur}.

We would like now to  comment  on the decoupling properties of heavy degrees of freedom for the renormalized or physical parameters \cite{appel}. As it is well known, the physical coupling constant does exhibit  these decoupling properties, thanks to the mass-dependence of the $RGE(M)$ equation, as given for instance by $\Delta g_M(g_M,M)$ in Eq.~(\ref{dec}). This is  not anymore the case for the  renormalized coupling constant in the $DR+MS$ scheme. These decoupling properties can of course be recovered explicitly in this scheme in terms of an effective local Lagrangian by considering only  light active degrees of freedom below heavy mass threshold \cite{collins}. This effective Lagrangian should thus incorporate, in principle, additional contributions of order $Q^2/m_{th}^2$, where $m_{th}$ is the  mass threshold. {\it This procedure is, however, not at all a necessity from a theoretical point of view}. 

The dependence of these coupling constants on the renormalization point $M$ on the one hand, and on the  regularization scale $\eta$ on the other hand, corresponds to two different properties. The first one is a property of the physical state realized in Nature, the second one is a property of the  local Lagrangian we start from. The absence of explicit decoupling for the renormalized parameters in the $DR+MS$ scheme is thus just a property of the local Lagrangian we start from and should not be corrected for since it is indeed correct! The scaling properties of the  renormalized parameters should refer to all the degrees of freedom present in the Lagrangian since they reflect its ultraviolet properties: these ones are independent of the (finite) masses of these elementary degrees of freedom. The dependence of $g_{R}$ on a dimensionless  scale  $\eta$ prevents also any attempts to correct it for mass thresholds.
The relationship between these two different behaviors, for $g_{R}(\eta)$ and $g_M(M)$, can be clearly seen from the decoupling of the  $\eta$-dependent  terms to the $M$-dependent ones in the calculation of the physical coupling constant, as shown in Subsec.~\ref{decoup}. The contribution $\Delta g_M$ does exhibit the decoupling of heavy degrees of freedom - $g_M$ being independent of $M$ for $m \gg M$ - while $\Delta g_{R}$ does not.

Our derivation of the $RG$ equations, in terms of the two complementary scales $\eta$ and $M$ associated  with the regularization procedure and with the renormalization scheme respectively, has also important consequences, as far as the unification of the coupling constants of the Standard Model is concerned \cite{ama}. The discussion on this possible unification is usually made in terms of the renormalized coupling constant $g_{R}$ in the $DR+MS$ scheme. As we argued in this study, the relevant regularization scale $\eta$ should not have  any interpretation in terms of a physical energy scale. It is also dimensionless. This coupling constant is therefore not the right quantity to look at in order to investigate the physical behavior of the Standard Model at high energy. This question should be addressed in terms of the physical coupling constant $g_M(M)$. In this case, the $RG$ equation $RGE(M)$ should be considered, in terms of the physical scale $M$. This $RG$ equation is mass dependent, and the question of the unification of the coupling constants of the Standard Model at very high energies should thus be investigated in the spirit of Ref.~\cite{bin}. As argued in this reference, mass corrections around the grand unification scale prevent any possibility for the coupling constants to cross at a single point. They should just converge and merge to a single coupling constant beyond this grand unification scale. Therefore, and in the absence of knowledge of the grand unified Lagrangian beyond this scale, nothing can be said quantitatively about the possible unification of interactions at this scale. This is in complete agreement with the way the scale of new physics, determined by $\Lambda_{NP}$ in (\ref{Leff}), can be fixed, as  mentioned there.

The determination of the scale dependence of the various parameters which enter into the calculation of a physical process  should rely on physical considerations. This is done in our study by always relating them to physical observables. This enables us  to calculate the dependence of the bare and renormalized coupling constants and masses on the regularization scale $\eta$.  We show  that using a coherent extension of the bare, renormalized and physical parameters at $\epsilon \neq 0$, the bare parameters of the local Lagrangian we start from should be regularization scale dependent, contrarily to what is usually assumed. 

We found however in a two-loop calculation within the $\phi^4$ theory, in the $DR+MS$ scheme, that the coefficients $\beta$ and $\gamma_m$ of the $RGE(\eta)$ equation are scale-independent and identical, at that order, to the ones already known using scale-independent bare parameters. While this result is gratifying in the sense that it does not rule out previous calculations in this particular model, it should be checked in a more elaborate calculation.

As we argued in this study, the calculation of the scale dependence of the renormalized parameters within the $DR+MS$ scheme can be done without the need to reinforce by hand the decoupling of heavy degrees of freedom. This is a direct consequence of the different intrinsic nature of the dimensionless regularization scale $\eta$ and the dimensionful renormalization point $M$. This may lead to sizeable corrections to the calculation of physical observables in the Standard Model. The difference in the intrinsic nature  of both scales should lead also to important consequences as far as the interpretation of the running of the bare,  renormalized or physical parameters is concerned.  This is particularly true with the identification of the relevant regularization scale as a dimensionless variable. 

\section*{Acknowledgments}
We would like to acknowledge enlightning discussions with P. Grang\'e and E. Werner.


\end{document}